\documentclass[aps,floats]{revtex4}
\usepackage{amsmath,amssymb}
\usepackage{graphicx,epsfig}

\begin{document}
\bibliographystyle {plain}

\def\oppropto{\mathop{\propto}} 
\def\opsimeq{\mathop{\simeq}}
\def\opoverderline{\mathop{\overline}}
\def\operarrow{\mathop{\longrightarrow}}
\def\opsim{\mathop{\sim}}

\def\fig#1#2{\includegraphics[height=#1]{#2}}
\def\figx#1#2{\includegraphics[width=#1]{#2}}


\title{  Numerical study of the directed polymer in a $1+3$ dimensional
random medium }

\author{ C\'ecile Monthus and Thomas Garel}
 \affiliation{Service de Physique Th\'{e}orique, CEA/DSM/SPhT\\
Unit\'e de recherche associ\'ee au CNRS\\
91191 Gif-sur-Yvette cedex, France}

\begin{abstract}

\bigskip

The directed polymer in a $1+3$ dimensional
 random medium is known to present a disorder-induced phase
transition. For a polymer of length $L$, the high temperature
phase is characterized by a diffusive behavior for the end-point
displacement $R^2 \sim L$ and by free-energy fluctuations of order
$\Delta F(L) \sim O(1)$. The low-temperature phase is characterized by
an anomalous wandering exponent  $R^2/L \sim L^{\omega}$ and
by free-energy fluctuations of order $\Delta F(L) \sim L^{\omega}$
where $\omega \sim 0.18$. In this paper, we first study the scaling
behavior of various properties to localize the critical temperature
$T_c$. Our results concerning $R^2/L$ and $\Delta F(L)$ point towards
$0.76 < T_c \leq T_2=0.79$, so our conclusion is that  $T_c$ is equal
or very close to the upper bound $T_2$ derived by Derrida and
coworkers ($T_2$ corresponds to the temperature above which the ratio
$\overline{Z_L^2}/(\overline{Z_L})^2$ remains finite as $L \to
\infty$). We then present histograms for the free-energy, energy and
entropy over disorder samples. For $T \gg T_c$, the free-energy
distribution is found to be Gaussian. For $T \ll T_c$, the free-energy
distribution coincides with the ground state energy distribution, in
agreement with the zero-temperature fixed point picture. Moreover the
entropy fluctuations are of order $\Delta S \sim L^{1/2}$ and follow a
Gaussian distribution, in agreement with the droplet predictions,
where the free-energy term $\Delta F \sim L^{\omega}$ is a near
cancellation of energy and entropy contributions of order $L^{1/2}$.

\end{abstract}

\maketitle


\section{Introduction }

The model of a directed polymer in a random medium 
 plays the role of a spin glass toy model in the field of disordered systems
\cite{Hal_Zha,Der_Spo,Der,Mez,Fis_Hus}.  At low
temperature, there exists a disorder dominated phase, where the 
order parameter is an `overlap' \cite{Der_Spo,Mez,Car_Hu,Com}.
In finite dimensions, a scaling droplet theory was proposed
 \cite{Fis_Hus,Hwa_Fis},
in direct correspondence with the droplet
 theory of spin-glasses \cite{Fis_Hus_SG},
whereas in the mean-field version of the model on the Cayley,
a freezing transition very similar to the one occurring
in the Random Energy Model was found \cite{Der_Spo}.
The phase diagram as a function of space dimension $d$ is the
following \cite{Hal_Zha}. In dimension $d \leq 2$, there is no free phase,
i.e. any initial disorder drives the polymer into the strong disorder phase,
whereas for $d>2$, 
there exists a phase transition between
the low temperature disorder dominated phase
and a free phase at high temperature  \cite{Imb_Spe,Coo_Der}.
 This phase transition 
has been studied exactly on a Cayley tree \cite{Der_Spo}
and on hierarchical lattice \cite{Der_Gri}.
  In finite dimensions, bounds on the critical temperature
$T_c$ have been derived \cite{Coo_Der,Der_Gol,Der_Eva} :
$T_0(d) \le T_c \le T_2(d)$.
The upper bound $T_2(d)$ 
corresponds to the temperature above which the ratio
$\overline{Z_L^2}/(\overline{Z_L})^2$ remains finite as $L \to
\infty$. The lower bound $T_0$ corresponds to the temperature below which
the annealed entropy becomes negative.
On the Cayley tree, the critical temperature $T_c$ coincides
with $T_0$ \cite{Der_Spo}.
In finite dimensions, there is a debate to know
whether $T_c$ coincides or not with $T_2$.
Arguments in favor of the equality $T_c=T_2$ 
in finite dimensions $d$ where $\omega(d)>0$
can be found in \cite{DPcritidroplet},
whereas a new upper bound $T^*$ based on the entropy of random
walks was recently proposed in \cite{birkner}. 
As explained in detail in \cite{talkihp}, the debate actually
concerns the form of the negative tail of the free-energy distribution
in the high temperature phase. If this tail decays only exponentially,
as on the Cayley tree, then $T_c<T_2$ is possible; if it decays more
rapidly than a simple exponential one gets $T_c=T_2$.  

In this paper, we study numerically the scaling properties
of the directed polymer in dimension $1+d$ with $d=3$.
As for previous numerical studies of the freezing
transition in $d=3$, we are only aware of  ref.\cite{Der_Gol,Ki_Br_Mo},
which draw different conclusions for the specific heat exponent $\alpha$,
namely $\alpha  \simeq -0.1 \pm 0.1$ in \cite{Der_Gol}
and $\alpha =2-\nu \sim -2 \pm 0.7$ in \cite{Ki_Br_Mo}.
With our numerical data, we are not able to give
a reasonable quantitative estimate of this critical exponent (see below).
However, our results concerning the scaling of the transverse
displacement $R^2(L)$ and of the free-energy fluctuations
$\Delta F(L)$ point towards a critical temperature
$T_c$ which is equal or very close to the upper
bound $T_2$ discussed above.
We also present histograms of free-energy, energy and entropy
as a function of temperature, in order to better understand
the nature of the high and low temperature phases.

The paper is organized as follows.
In Sect \ref{model}, we define the model and give some numerical details.
In Sect. \ref{spatial}, we present our results concerning
the end-point fluctuations.
In Sect. \ref{thermodynamic}, we discuss the scaling behavior
and the probability distribution 
of the free-energy, the energy and the entropy as $T$ varies.
In Sect. \ref{fss}, we present the finite-size scaling analysis
of spatial and thermodynamic observables.
We summarize our conclusions in Sect. \ref{conclusion}.

\section{ Model and Numerical details}

\label{model}

In this paper, we present numerical results
for the random bond version
of the model defined by the recursion relation
on a cubic lattice in $d=3$
\begin{eqnarray}
\label{DP1}
Z_{L+1} (\vec r) =  \sum_{j=1}^{2d}
 e^{-\beta \epsilon_L(\vec r+\vec e_j,\vec r)} Z_{L} (\vec r+\vec e_j)
\label{transfer}
\end{eqnarray}
The bond energies $\epsilon_L(\vec r+\vec e_j,\vec r) $
are random independent variables drawn from the Gaussian
distribution 
\begin{eqnarray}
\rho (\epsilon) = \frac{1}{\sqrt{2\pi} } e^{- \frac{\epsilon^2}{2} }
\end{eqnarray}
The numerical values of the bounds discussed in the Introduction
are then given by \cite{Der_Gol}
\begin{equation}
T_0(d=3)=0.528.. \le T_c \le T_2(d=3)=0.790...
\label{derridabounds}
\end{equation}
In the previous numerical study of the same model \cite{Der_Gol},
a finite size scaling analysis with respect to the transverse direction
gave a critical temperature of order $T_c \sim 0.6$,
whereas our data presented below point towards
$T_c \sim T_2=0.79$. 

In the high temperature phase ($T \geq T_c$), the free-energy density
\begin{eqnarray}
f(T) \equiv -T  \lim_{L \to \infty} \frac{ \ln Z_L }{L}
\end{eqnarray}
is known to coincide with the annealed value \cite{Der}
\begin{eqnarray}
f(T)=f_{ann}(T)= -T \ln (2d) - \frac{1}{2 T}
 \ \ \  {\rm  for } \ \ \ \  T \geq T_c 
\label{fhigh}
\end{eqnarray}
As a consequence, the energy $e(T)$ and the specific heat $c(T)$ also coincide
with their annealed values $e_{ann}(T)=-1/T$ and $c_{ann}(T)=1/T^2$,
so that the regular parts of thermodynamic quanties can
be exactly substracted out.

The numerical results presented in the following
have been obtained from the direct iteration of 
the transfer matrix equation (\ref{transfer}).
In Sect. \ref{spatial} concerning the end-point transverse
fluctuations, we give results for lengths $L$ in the range
$30 \leq L \leq 60$, with $n_s(L)$ independent samples
between $n_s(L=30)= 35. 10^4$ and $n_s(L=60)= 23. 10^3$.
In Sect. \ref{thermodynamic} concerning thermodynamic quantities,
we have used similar values, as well as smaller sizes to have a bigger
range in the size $L$.
 For instance, the data presented on Fig. (\ref{fig4}) 
correspond to lengths $L=6, \ 12, \ 24, \ 48$ with respective
numbers $n_s(L)$ of independent samples of order
 $n_s(L)= 9 \cdot 10^6, 9 \cdot 10^6, 6 \cdot 10^5, 4\cdot 10^4$.

\section{ Study of spatial properties  }

\label{spatial}

\begin{figure}[htbp]
\includegraphics[height=6cm]{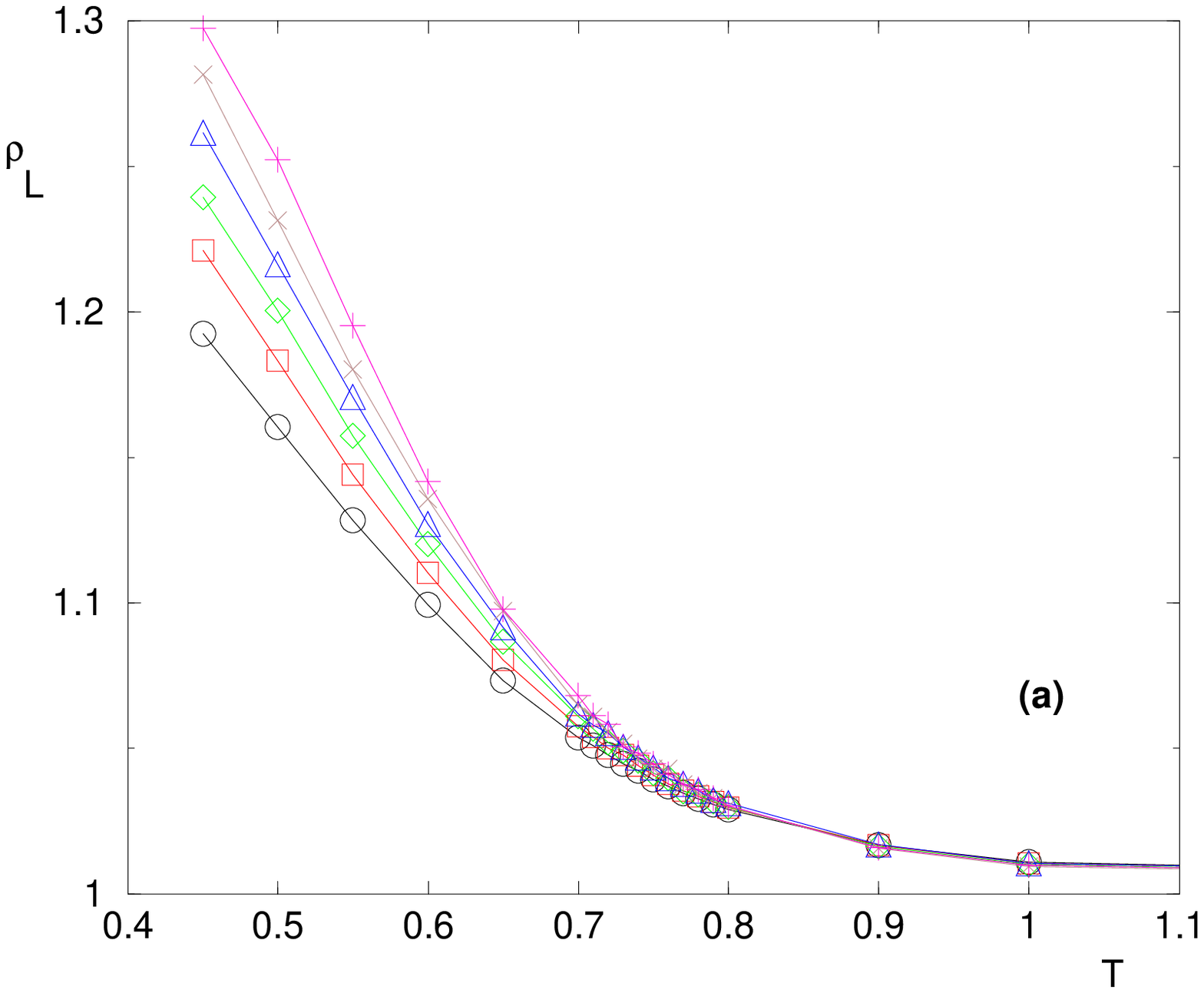}
\hspace{1cm}
\includegraphics[height=6cm]{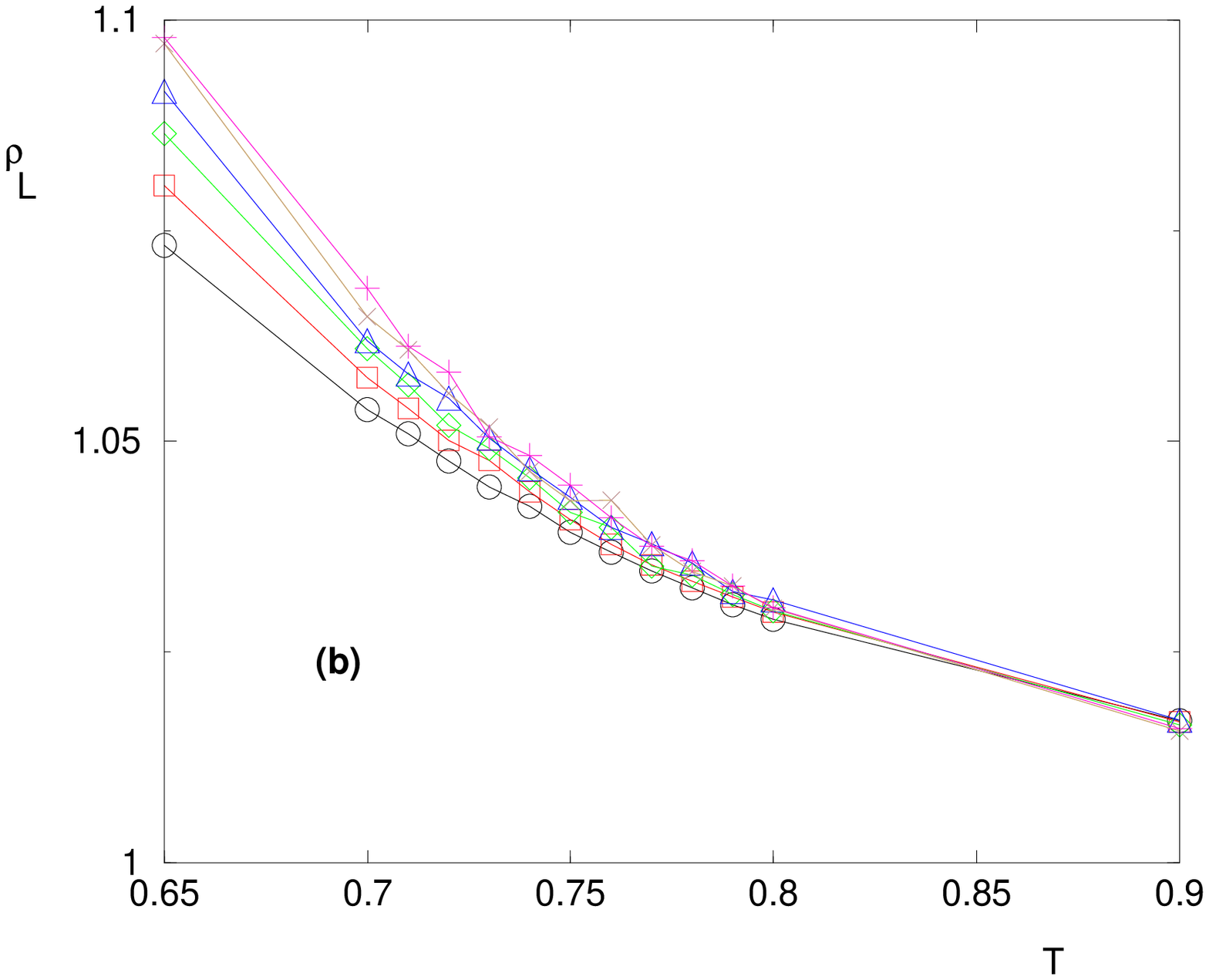}
\caption{(a) $\rho_L(T)=\overline{<R^2_L(T)>}/L $ as a
function of the temperature $T$ for different
$L=30$  $(\bigcirc)$ , $36$ $(\square)$  , $42$ $(\lozenge)$, $48$
$(\triangle)$, $54$ $(\times)$, $60$ $(+)$ (b) Zoom of the same data
in the critical region.}
\label{fig1}
\end{figure}

The phase transition corresponds to a change in the wandering exponent
$\zeta$ characterizing the transverse
 fluctuations $\overline{<R^2_T(L)>}   $ of the end-point of the
polymer \cite{Hal_Zha}
\begin{eqnarray}
\overline{ <R^2_T(L)> } \opsimeq_{L \to \infty} L \ \ {\rm for} \ \ T \geq T_c \\
\overline{ <R^2_T(L)> } \opsimeq_{L \to \infty} L^{2 \zeta} \ \ {\rm for} \ \ T < T_c
\end{eqnarray}
where, as usual, $<A>$ and $\overline{A}$ denote respectively thermal
and disorder averages. The exponent $\zeta \sim 0.59...$ characterizes
the zero-temperature fixed point  \cite{Hal_Zha}.
On Fig. (\ref{fig1}), we show the data for
$\rho_L(T)=\overline{<R^2_T(L)>} /L  $ as a function of the temperature $T$
for increasing lengths in the range $30 \leq L \leq 60$.
To localize the critical temperature $T_c$, we have tried to fit our
data according to the following form

\begin{eqnarray}
\frac{ \overline{ < R^2_T(L) >}}{L} =a_0(T) L^{\omega} +a_1(T)+ \frac{a_2(T)}{L}
\label{gyr}
\end{eqnarray}
where we have used the scaling relation $\omega=2 \zeta-1 \sim 0.186$.
The coefficient $a_0(T)$ is found to vanish around $T_c \simeq 0.78-0.79$.
A tentative finite-size scaling analysis is shown on Fig. (\ref{fig9}).

\section{ Study of thermodynamic properties}

\label{thermodynamic}

For a disordered sample $(i)$ of length $L$,
we note the free energy $F_T(L,i)$, the entropy $S_T(L,i)$ and the internal
energy $E_T(L,i)$. For each of these observables ($X=F,S,E$), we will
discuss the behavior of the averaged value $X_T^{av}(L)$, of the variance
 $\Delta X_T^{av}(L)$ and of the rescaled probability distribution
\begin{equation}
P_{(T,L)}(X)   \simeq  \frac{1}{ \Delta X_T(L)} \  
G \left( x= \frac{
X -X_T^{av}(L)}{ \Delta X_T(L) }  \right) 
\label{rescalinghisto}
\end{equation}

\subsection{ Delocalized phase at very high temperature $T \gg T_c$ } 

\begin{figure}[htbp]
\includegraphics[height=6cm]{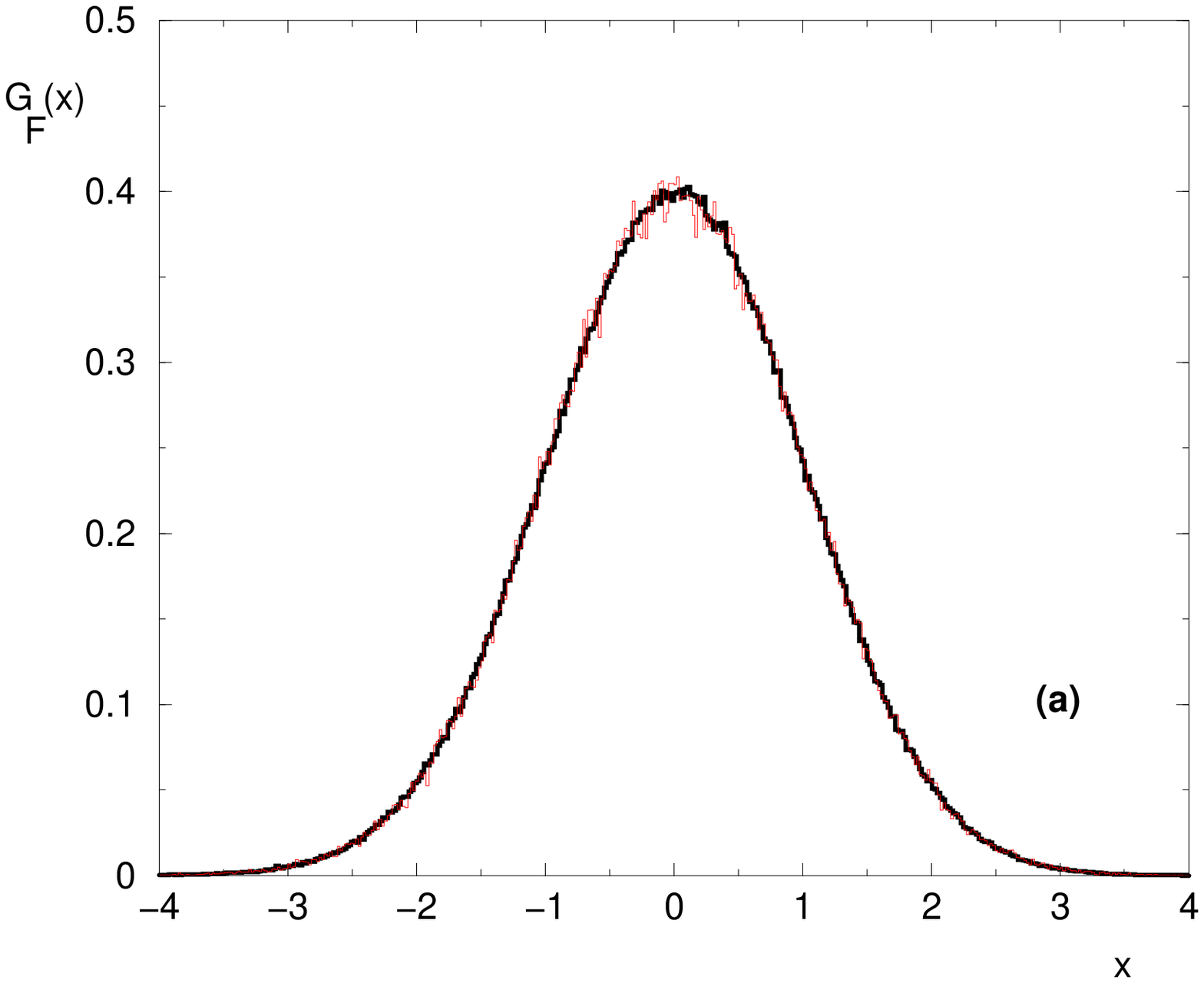}
\hspace{1cm}
\includegraphics[height=6cm]{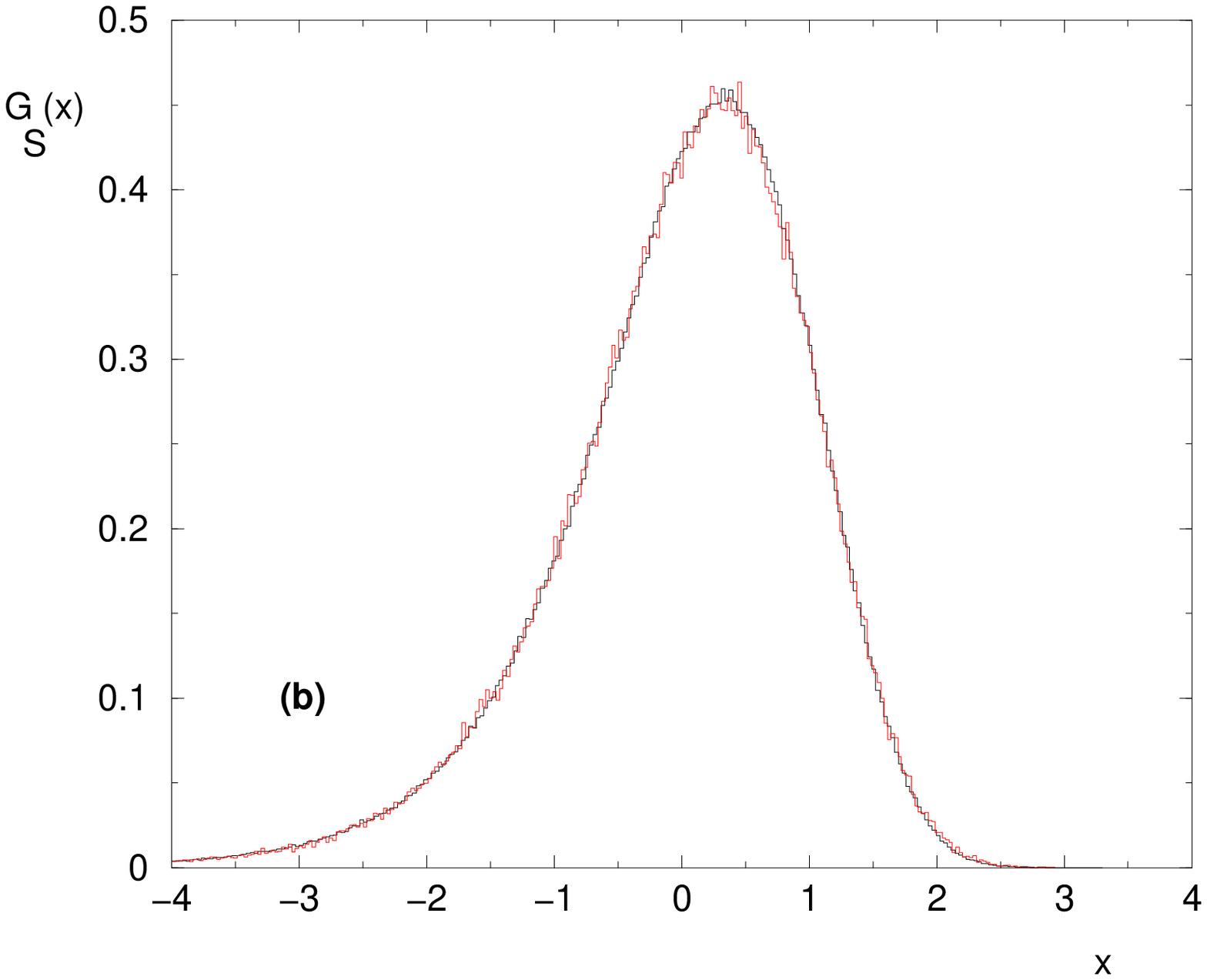}
\caption{ High temperature phase: (a) Gaussian rescaled probability
distribution $G_F(x)$ of the free energy at $T=2$ for $L=18$, $30$.
(b) Non-Gaussian rescaled probability distribution $G_S(x)$ 
of the entropy at $T=2$ for $L=18, \ 30$. }
\label{fig2}
\end{figure}

In the high temperature phase $T \gg T_c$, the free-energy $F_T(L,i)$,
the entropy $S_T(L,i)$ and the energy $E_T(L,i)$
of a sample $(i)$ of length $L$ are expected to be 
\begin{eqnarray}
F_T(L,i)=L f_{ann} + a_i \\
S_T(L,i)=L s_{ann} + b_i \\
E_T(L,i)=L e_{ann} + c_i
\label{fdelocalized}
\end{eqnarray}
where $a_i,b_i,c_i$ are random variables of order $O(1)$. The rescaled
distribution (\ref{rescalinghisto}) of the free energy is found to be
Gaussian (see Fig. \ref{fig2} (a)), whereas the rescaled probability
distribution of the entropy is asymmetric  (see Fig. \ref{fig2} (b)).

\subsection{ Localized phase at very low temperature $T \ll T_c$ } 

\begin{figure}[htbp]
\includegraphics[height=6cm]{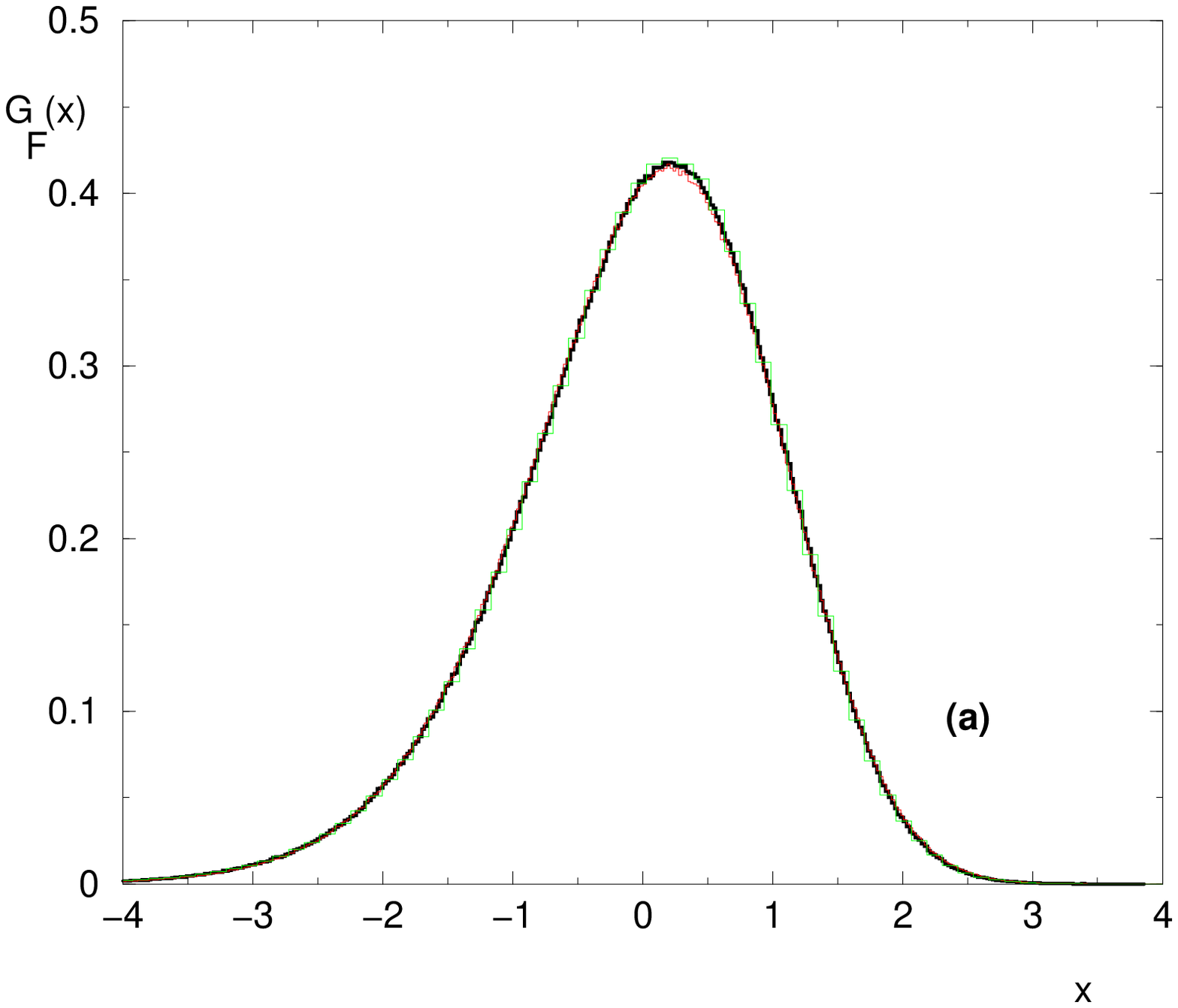}
\hspace{1cm}
\includegraphics[height=6cm]{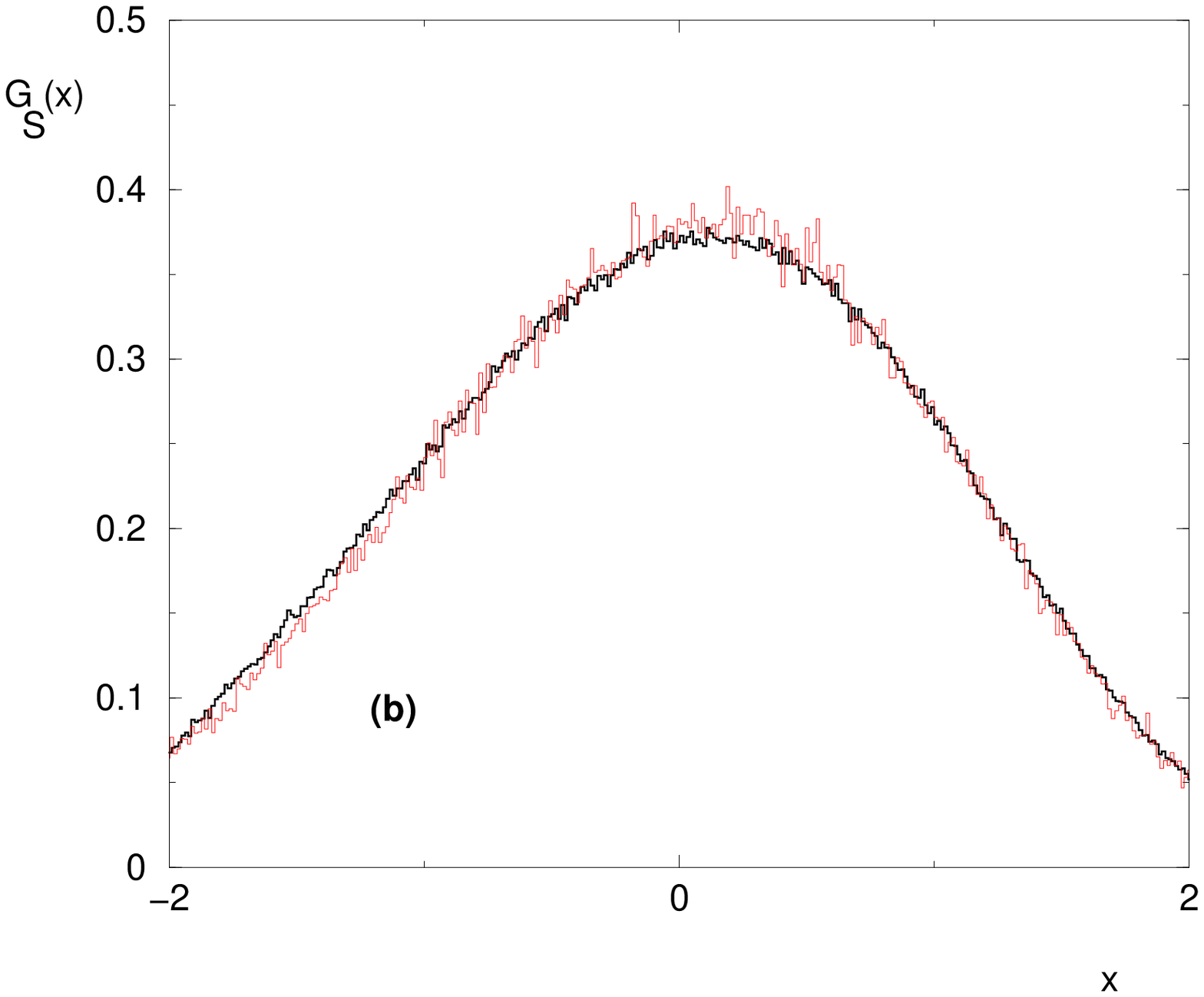}
\caption{ (a) $T=0$ fixed
point: comparison of the rescaled distributions of
the ground state energy $G_{E_0}(x)$ and of the free
energy $G_F(x)$ at $T=0.25, \ 0.4$ for $L=12$ (b) The rescaled entropy
distribution $G_S(x)$ for $T=0.25$ and $L=18,\ 30$ is found to be
Gaussian in agreement with the droplet picture.} 
\label{fig3}
\end{figure}

Let us first recall some theoretical expectations.
At $T=0$, the ground state energy of a sample $(i)$ of size $L$
takes the scaling form
\begin{eqnarray}
E_0(L,i)=L e_0 + L^{\omega} u_i
\label{ground}
\end{eqnarray}
where $u_i$ is a random variable of order $O(1)$.
As a consequence, both the correction to extensivity 
and the fluctuations involve the same exponent 
 $\omega \sim 0.186$
\begin{eqnarray}
\overline{E_0(L,i)}-L e_0 && = L^{\omega} \overline{u_i} \\
\Delta E_0(L) && = L^{\omega}
 \left( \overline{u_i^2} - (\overline{u_i})^2  \right)^{1/2}
\label{singleground}
\end{eqnarray}
Numerical results on the statistical properties of the ground state
and low energy excitations are discussed in our recent work
\cite{DPexcita}. According 
to the droplet theory, the whole low temperature phase $0<T<T_c$ 
is governed by a zero-temperature fixed point. 
However, many subtleties arise because the temperature
is actually `dangerously irrelevant'. 
The main conclusions of the droplet analysis \cite{Fis_Hus}
can be summarized as follows.
The scaling (\ref{ground}) translates into the
statistical properties of the free energy, provided one introduces
a correlation length $\xi(T)$ to rescale the length $L$
\begin{eqnarray}
F_T (L,i)-L f_{ann} = \left( \frac{L}{\xi(T) } \right) +
 f_1(T) \left( \frac{L}{\xi(T) } \right)^{\omega} u_i
\label{freelow}
\end{eqnarray}
where $u_i$ is a random variable of order $O(1)$.

For $T$ low enough ($T \leq 0.4$), we find that the width $\Delta
F(L,T)$ indeed scales as $L^{0.18}$. Moreover the rescaled
distribution (\ref{rescalinghisto}) coincides with
the rescaled distribution of the ground state energy at $T=0$ (see Fig
\ref{fig3}(a)).

However, within the droplet theory \cite{Fis_Hus},
the random part of order $L^{\omega}$ of the free-energy
(Eq. \ref{freelow})
is expected to be a near cancellation of energy and
entropy random contributions of order $L^{1/2}$.
The argument is that the energy and entropy are dominated by small
scale contributions 
of random sign \cite{Fis_Hus}, whereas the free energy is optimized
on the coarse-grained scale $\xi(T)$.
These predictions for the energy and 
entropy have been numerically checked in $d=1$
and $d=2$ \cite{Fis_Hus,Wa_Ha_Sc}. 
Moreover, the droplet argument \cite{Fis_Hus} suggests that the
entropy can be written as
\begin{eqnarray}
 S_{T} (L,i)-L s_{ann}  = s_0(T) L +s_1(T) L^{1/2} v_i+...
\label{entropylow}
\end{eqnarray}
where $v_i$ is Gaussian, because the entropy
 is dominated
by a large number of independent local excitations.
We show on Fig. \ref{fig3} (b) that the rescaled distribution is 
in good agreement with this argument.

Then from the thermodynamic relation $F=E-T S$,
 the energy is expected
to have a more complicated form involving terms of order $L$, $L^{1/2}$
and $L^{\omega}$.

\subsection{ Critical region}

\subsubsection{ Free-energy scaling at $T_c$  }

\begin{figure}[htbp]
\includegraphics[height=6cm]{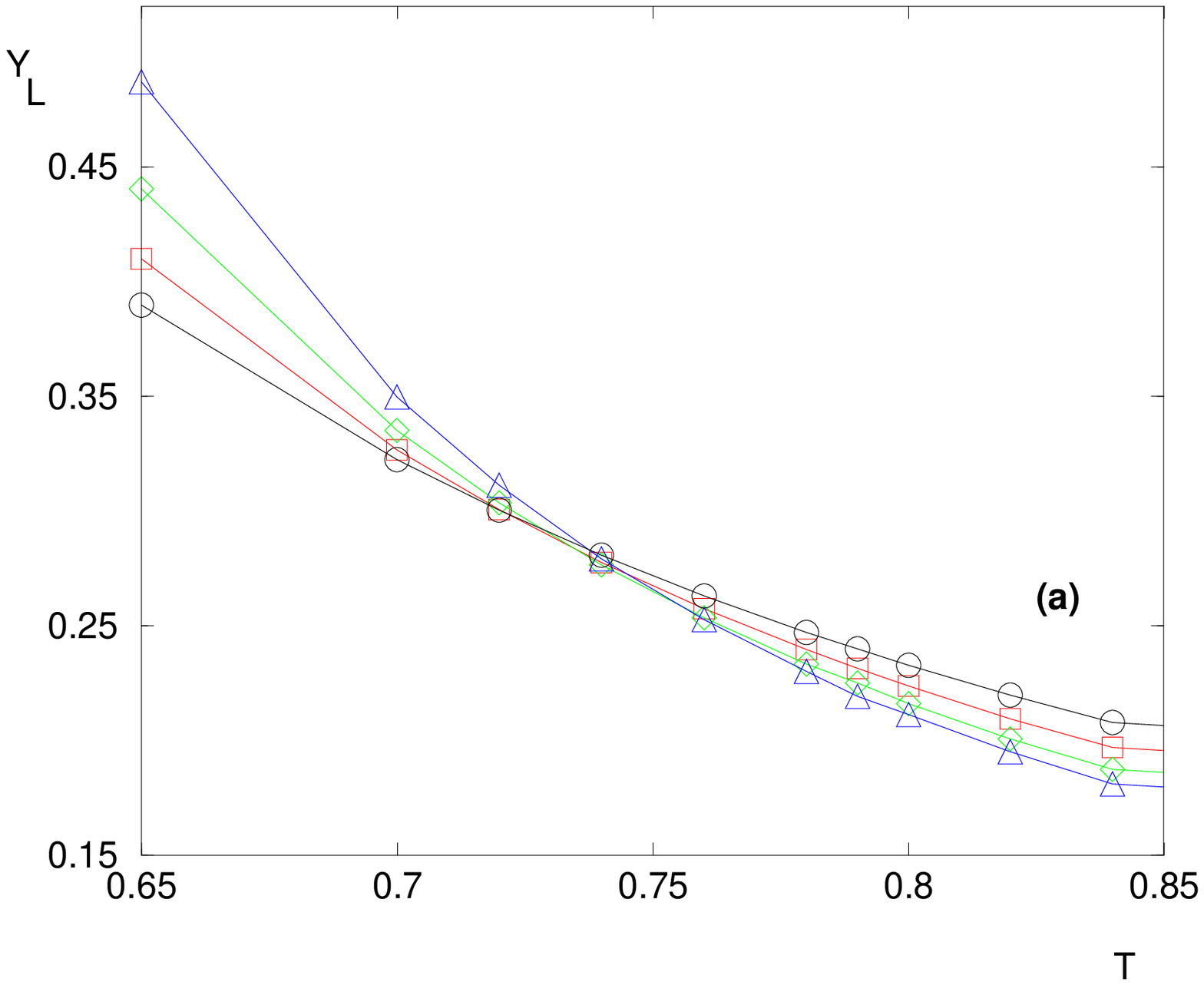}
\hspace{1cm}
\includegraphics[height=6cm]{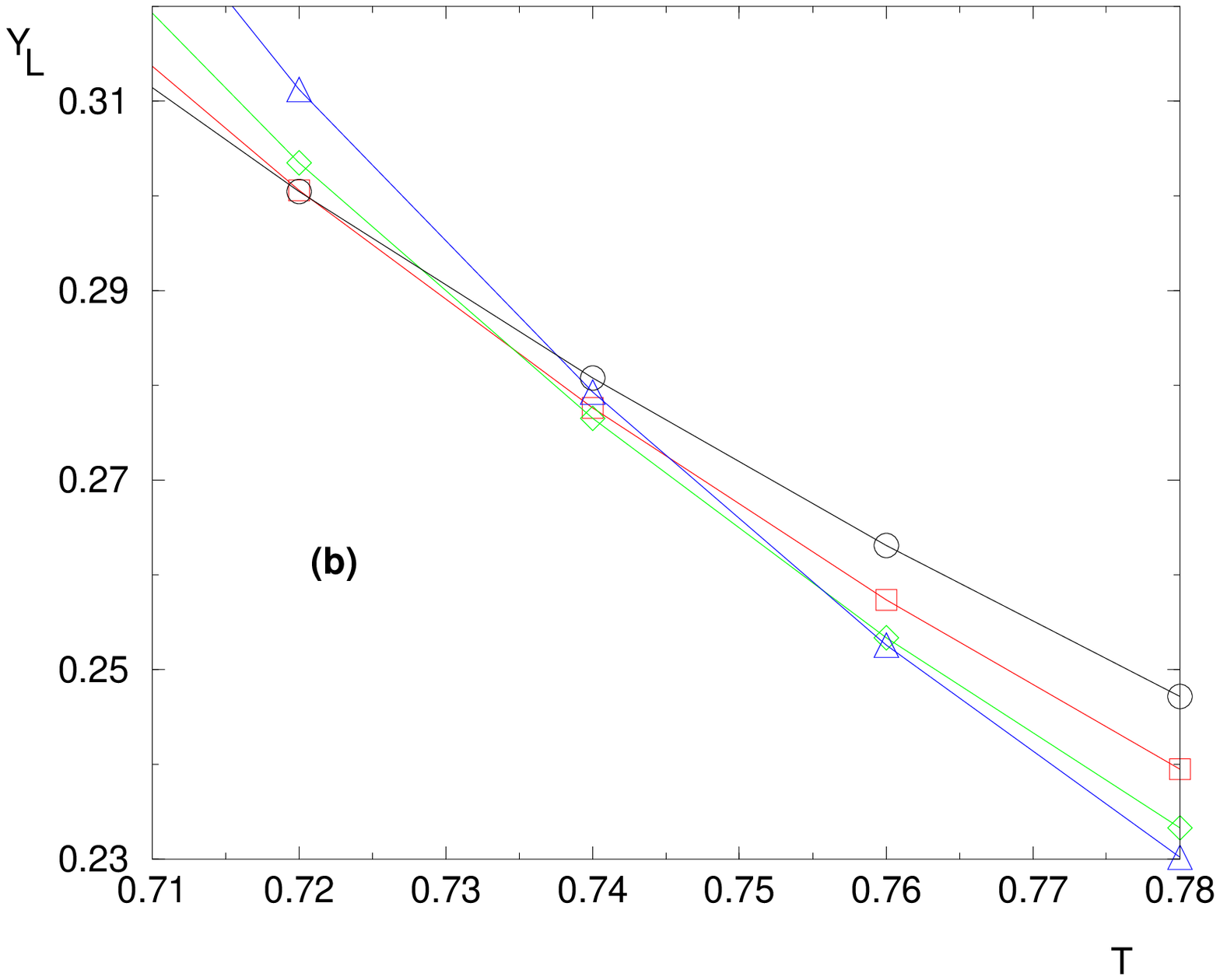}
\caption{ (a) Plot of $ Y_L(T)=\frac{F_{T}^{av} (L)-L
f_{ann}}{(\ln L)^{1/2}}$ for $L=6$ $(\bigcirc)$, $12$ $(\square)$,
$24$ $(\lozenge)$, $48$ $(\triangle)$ (b) Zoom around the
pseudo-crossing point, showing a systematic shift with $L$ : 
the curves for $L=6$ and $L=12$ cross around $T \sim 0.72$,
whereas the curves for $L=24$ and $L=48$ cross around $T \sim 0.755$.} 
\label{fig4}
\end{figure}

At $T_c$, the wandering exponent $\zeta$ is expected
to be exactly $\zeta_c=1/2$ \cite{Do_Ko}, and thus the scaling
relation $\omega_c=2 \zeta_c-1$  between exponents \cite{Hus_Hen}
yields $\omega_c=0$.
Previous numerical studies and arguments \cite{Ki_Br_Mo,Fo_Ta} 
have concluded that the free-energy is
 logarithmic at $T_c$ with exponent $(1/2)$
\begin{eqnarray}
 F_{T_c} (L,i)-L f_{ann}  = (\ln L)^{1/2} w_i
\label{fcriti}
\end{eqnarray}
where $w_i$ is a random variable of order $O(1)$. We show on
Fig. \ref{fig4}(a) that the rescaled variable $
Y_L(T)=\frac{F_{T}^{av} (L)-L f_{ann}}{(\ln L)^{1/2}}$ increases with
$L$ in the low temperature phase, and decreases with $L$ in the high
temperature phase. In the critical region, there is a systematic shift
in the crossing points of curves when $L$ increases (see the zoom
of Fig. \ref{fig4}(b)). Since our numerical data are not sufficient
to analyse quantitatively the $L$-dependence of this shift,
we simply conclude that the thermodynamic critical temperature $T_c$
is in the region $T_c \geq 0.76$,
and is thus very close to the exact upper bound 
 $ T_2=0.79$  (Eq. \ref{derridabounds}).

\subsubsection{ Rescaled distributions of free energy, energy and
entropy in the critical region}

\begin{figure}[htbp]
\includegraphics[height=6cm]{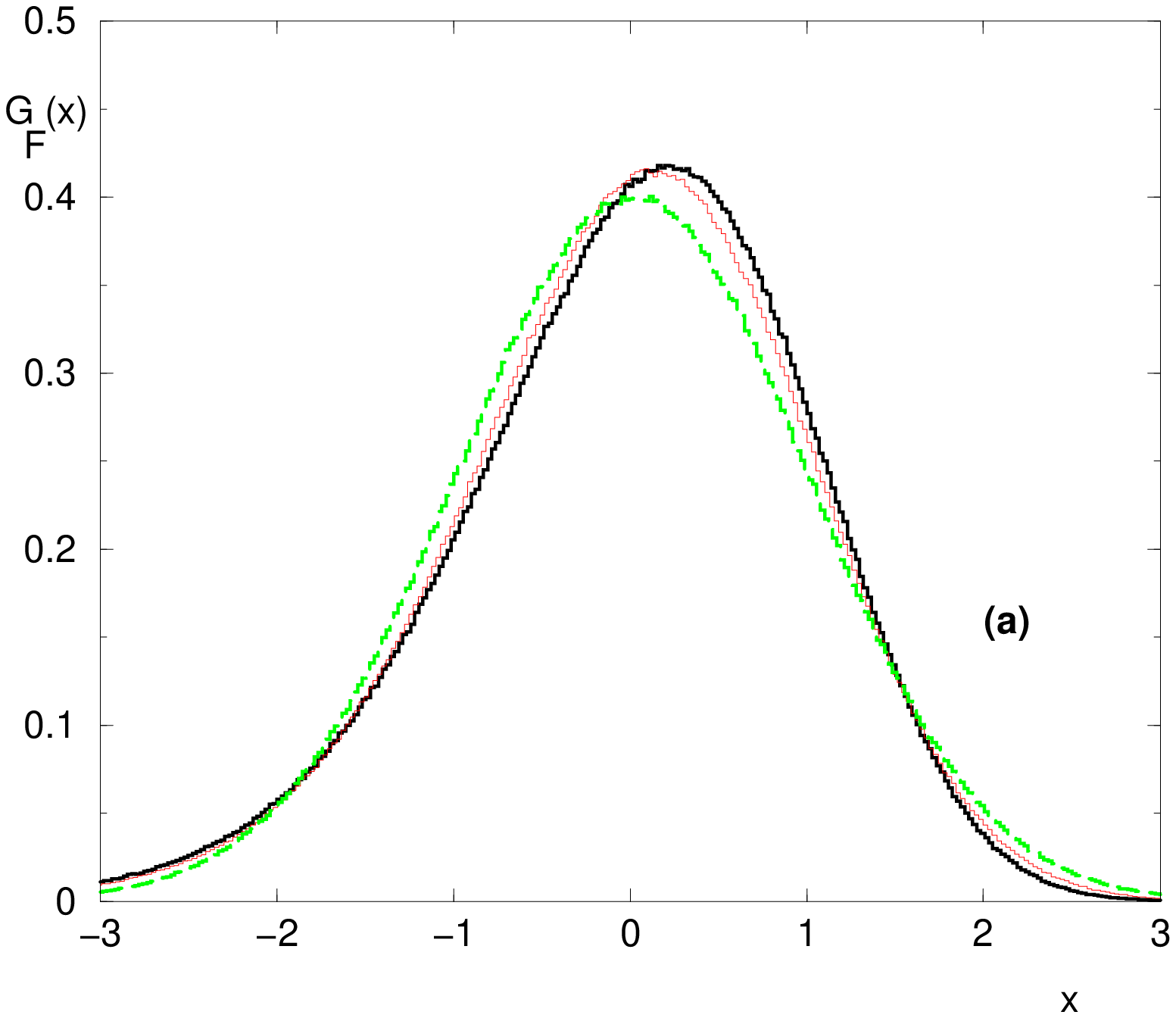}
\hspace{1cm}
\includegraphics[height=6cm]{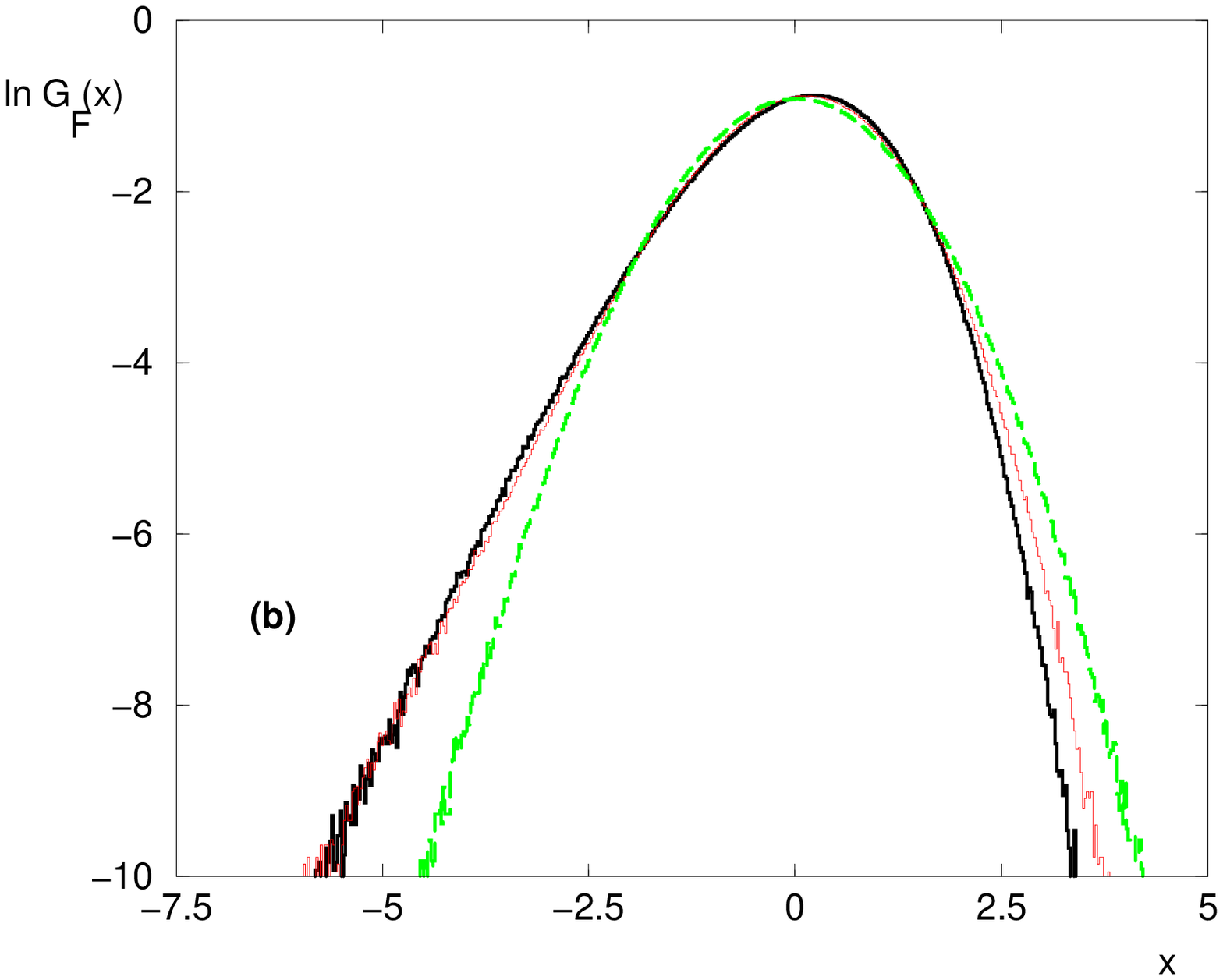}
\caption{ Free energy distribution : (a)  Rescaled probability distribution
$G_F(x)$ of the free-energy for $L=12$ in the low temperature phase
$T=0.25$ ( thick line)),
in the high temperature phase $T=2$ (Gaussian, long dashed line) and in the
critical region  $T=0.79$ (thin line).
(b)  Logarithmic plot of the same data: in the critical region, the
negative tail is very close to the low-temperature tail and far from
 the Gaussian high temperature tail. 
 }
\label{fig2bis}
\end{figure}

\begin{figure}[htbp]
\includegraphics[height=6cm]{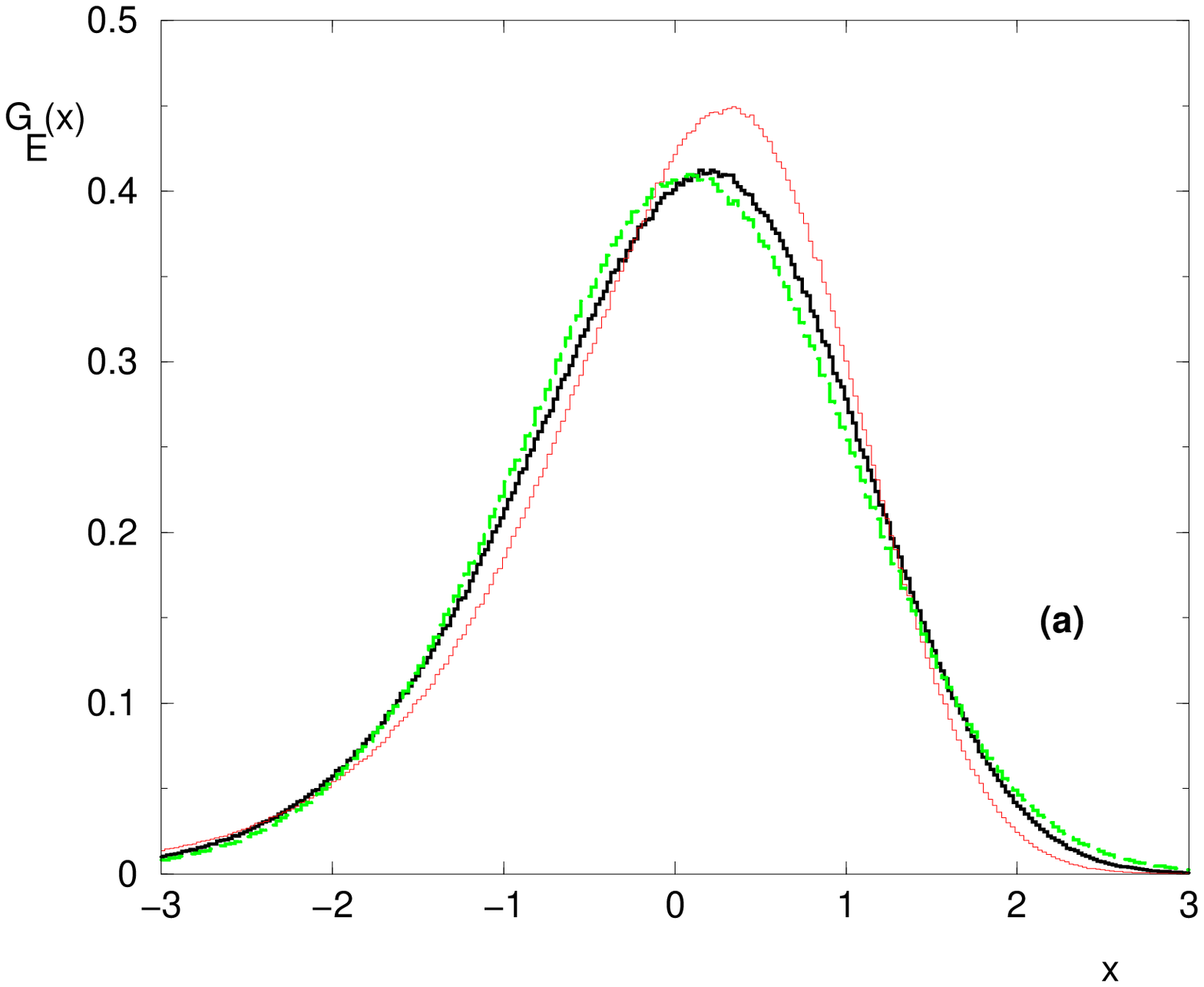}
\hspace{1cm}
\includegraphics[height=6cm]{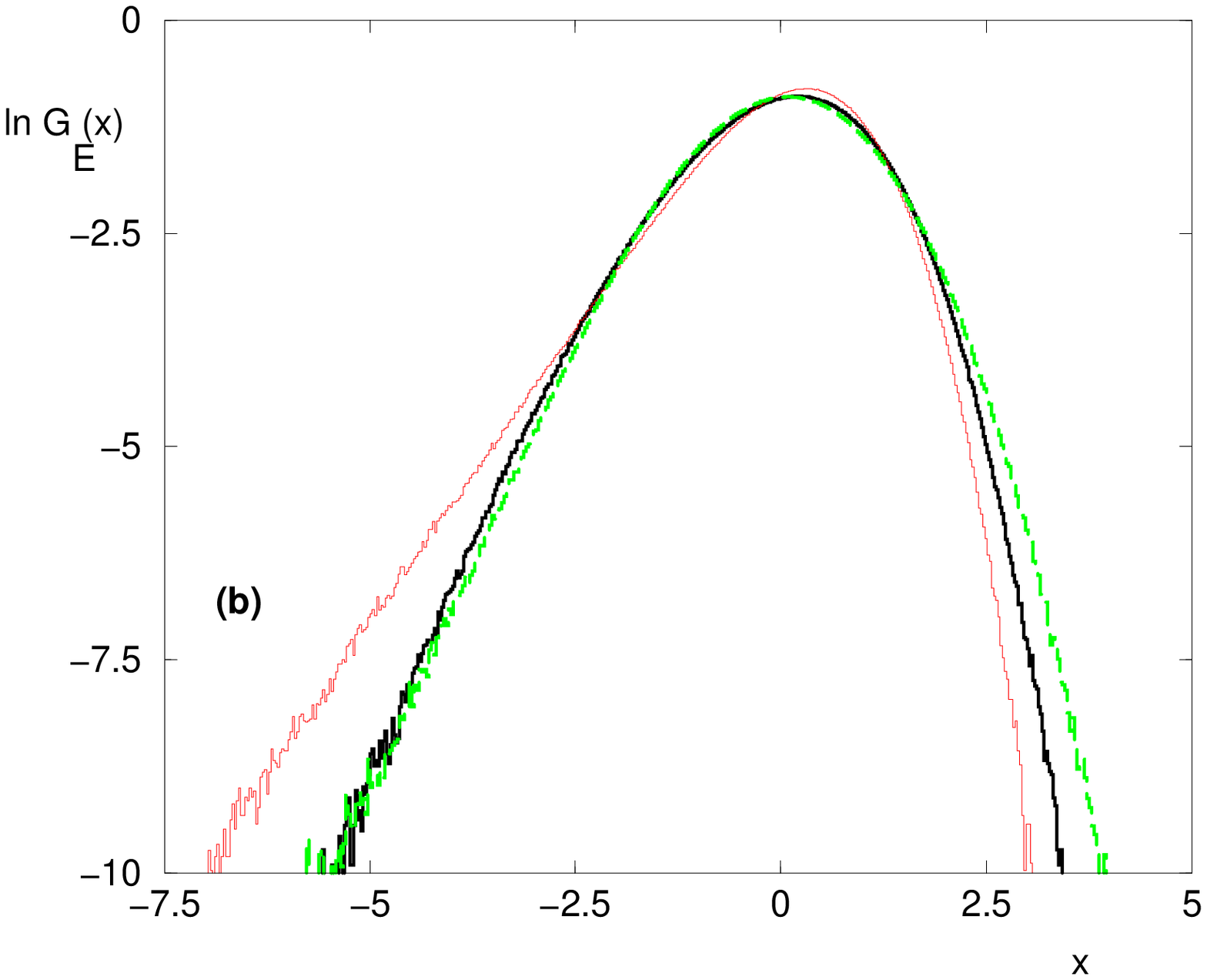}
\caption{(a)  Rescaled
energy distributions for $L=12$ in the low temperature phase
$T=0.25$ (thick line),
in the high temperature phase $T=2$ (long dashed line) and in the
critical region $T=0.79$ (thin line), where the distribution is
maximally peaked (b)  Logarithmic plot of the same data
:  the slower decay in the negative tail is obtained in the critical region
and seems to correspond to an exponential decay
$\ln G_{E}(x) \sim - \vert x \vert  $ near $T_c$  .}
\label{fig5}
\end{figure}

\begin{figure}[htbp]
\includegraphics[height=6cm]{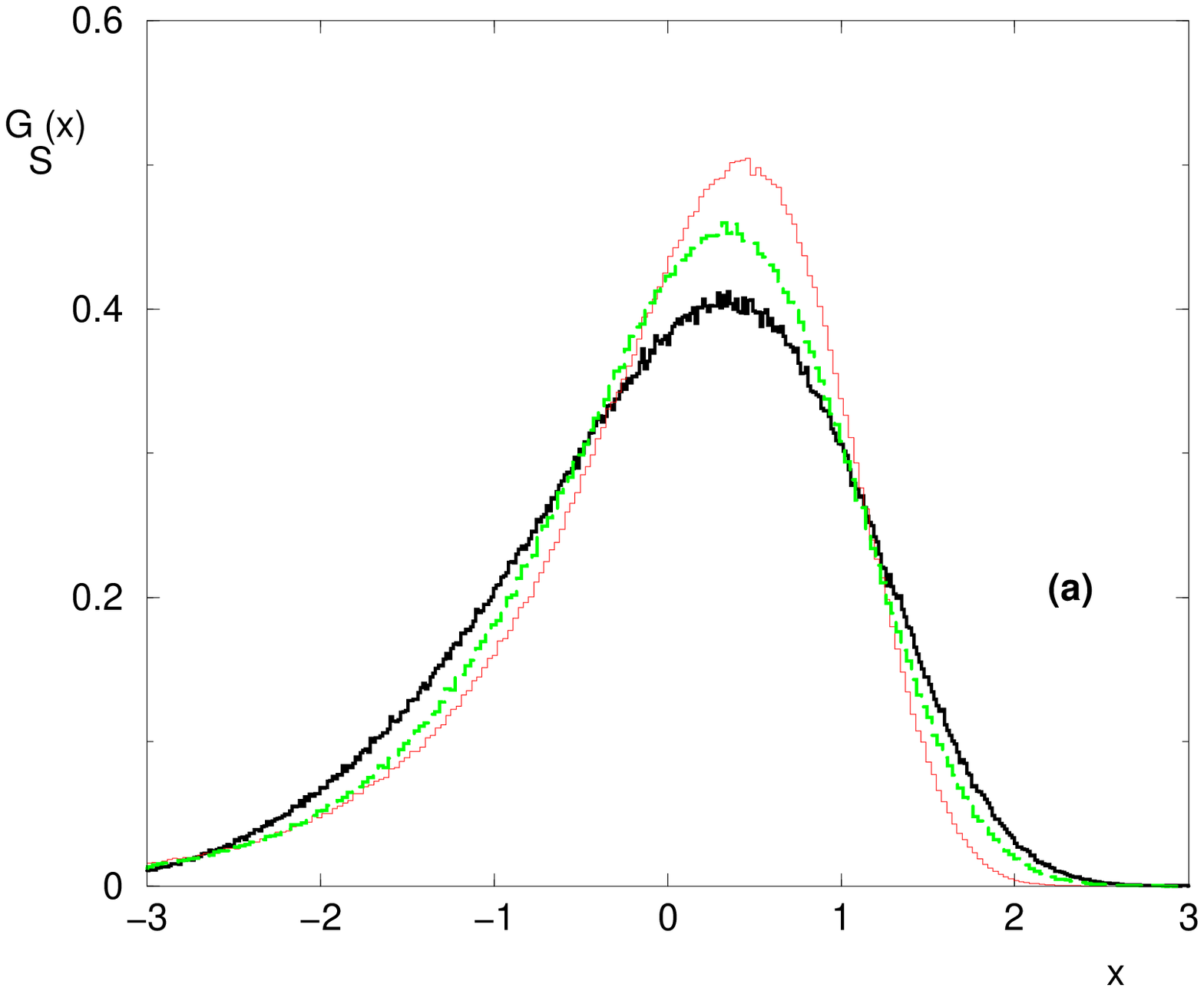}
\hspace{1cm}
\includegraphics[height=6cm]{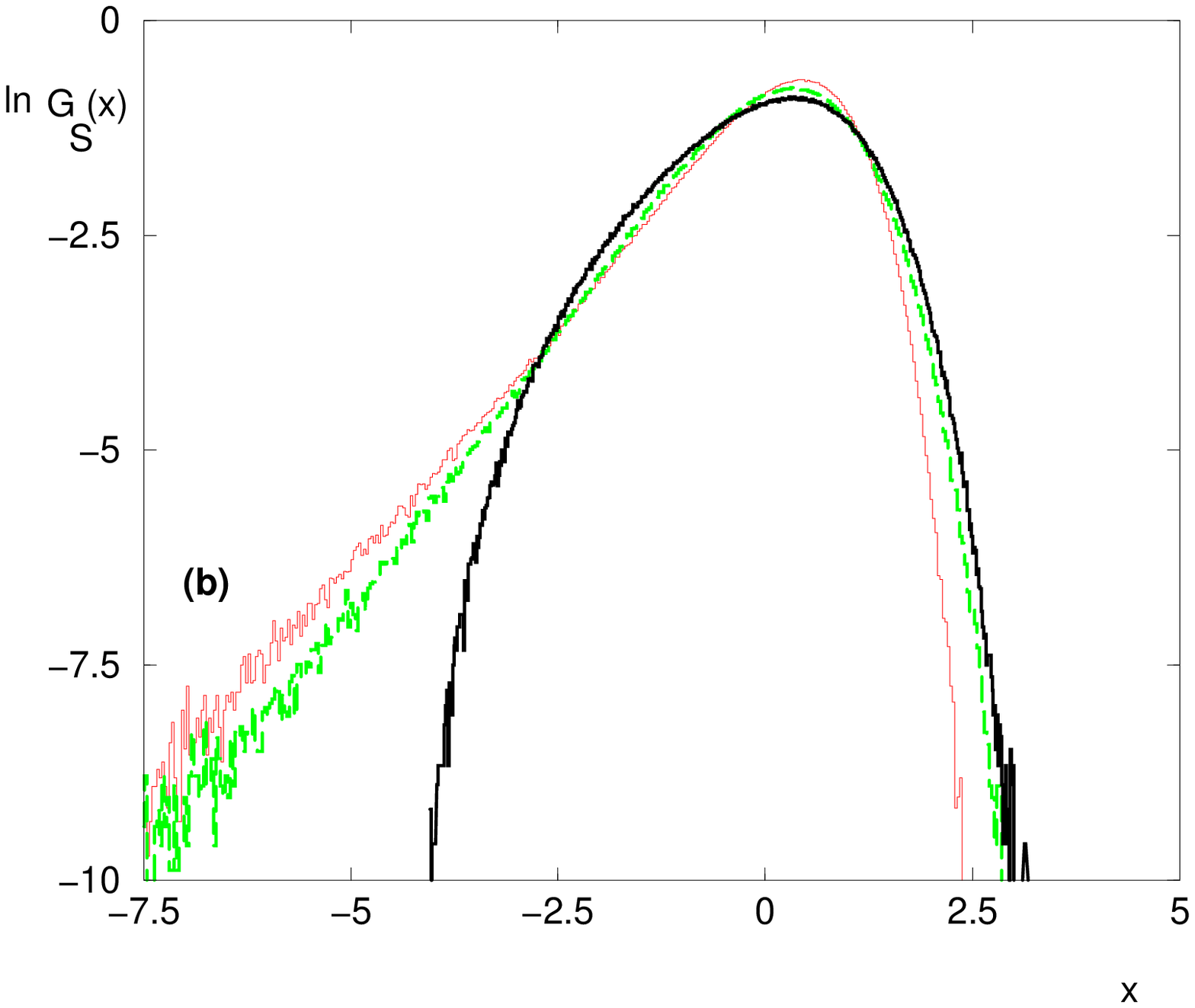}
\caption{ (a)  Rescaled
entropy distributions for $L=18$ in the low temperature phase
$T=0.4$ (thick line),
in the high temperature phase $T=2$ (long dashed line) and in the
critical region 
$T=0.79$ (thin line), where the distribution is maximally peaked
(b)  Logarithmic plot of the same data
: the slower decay in the negative tail is obtained in the critical region
and seems to correspond to an exponential decay
$\ln G_{S}(x) \sim - \vert x \vert  $ near $T_c$ .}
\label{fig6}
\end{figure}

In this section, we compare the rescaled probability distributions
of thermodynamic observables in the critical region 
with respect to the low temperature 
and high temperature corresponding results.

The results for the free-energy are shown on Figs. \ref{fig2bis}
focusing on the bulk (a) and on the tails (b). 
The negative tail of this probability distribution 
is of particular interest, since it is linked to the location
of the critical temperature $T_c$ with respect to the upper bound
$T_2$ as explained in details in
\cite{talkihp}.  If this tail is exponential $G_F(x) \sim e^{ - a
\vert x \vert }$ for $x \to -\infty$
 as on the Cayley tree \cite{Der_Spo},
then $T_c<T_2$, whereas if $G_F(x)$ decays more
 rapidly than exponentially, then $T_c=T_2$.
For instance, in the mean-field Sherrington-Kirkpatrick model of 
spin-glasses, one has $T_c=T_2$ and the distribution of the free-energy
fluctuations for $T>T_c$ is known to be Gaussian 
\cite{Aiz_Leb_Rue,Com_Nev}. 
Here we obtain numerically
that $G_F(x)$ is Gaussian in the high temperature phase for $T \gg T_2$
 (see Fig. \ref{fig2} (a)). If this Gaussian distribution is valid
in the whole high temperature phase, then one has the equality
$T_c=T_2$. On the other hand, the strict inequality $T_c<T_2$
requires that the negative tail becomes exponential in a region above
$T_c$. Such a `transition' in the form of the free-energy distribution
within a high temperature phase is known to exist for instance
in the Random Energy Model \cite{Bo_Ku_Lo}. Here with our numerical results,
as $T$ grows from $T_2$ to $T \gg T_2$, we see a slow crossover
from the low-temperature tail towards the Gaussian high temperature tail
(see  Fig. \ref{fig2bis}). Since $G_F(x)$ decays in the low-temperature phase
as $e^{- b \vert x \vert^{\eta}}$
 with $\eta=1/(1-\omega) >1$ according to Zhang
argument \cite{Hal_Zha,DPcritidroplet,talkihp},
 we see no sign of an exponential decay with power $\eta=1$
at any temperature.

The rescaled probability distribution 
 for the energy and entropy are shown respectively
on Figs. \ref{fig5} and \ref{fig6}
focusing on the bulk (a) and on the tails (b). 
Concerning the bulk (see Figs  \ref{fig5} (a) and \ref{fig6} (a)),
both distributions have a non-monotonous behavior as $T$ varies,
and become maximally peaked in the critical region $0.7 < T<0.9$.
As for the negative tails (see Figs \ref{fig5} (b) and \ref{fig6} (b)),
the slower decay is obtained in the critical region
and seems to correspond to an exponential decay
$G_{E,S}(x) \sim e^{- c \vert x \vert }$.

\newpage

\section{ Tentative finite size scaling analysis on averaged
observables}

\label{fss}

\begin{figure}[htbp]
\includegraphics[height=6cm]{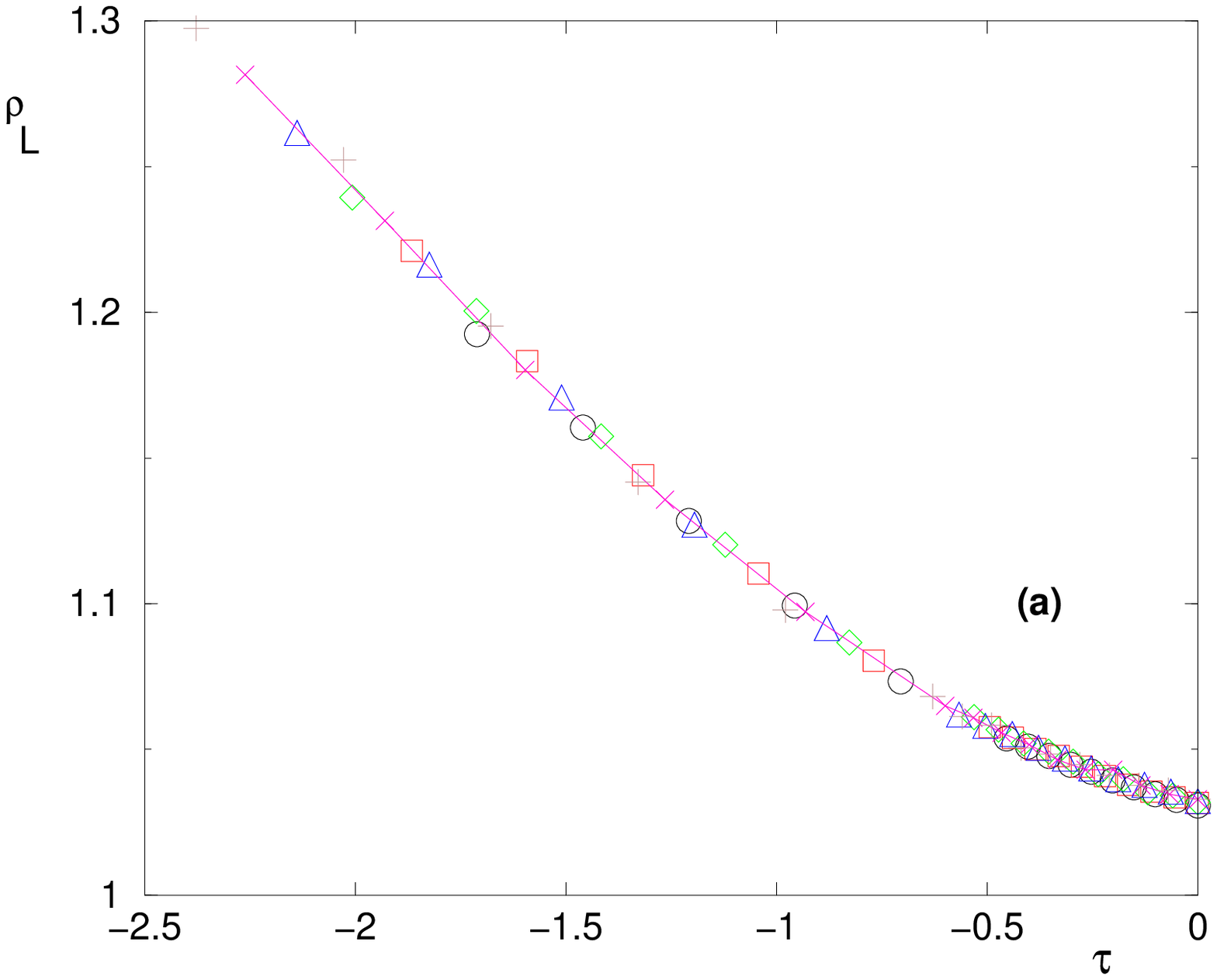}
\hspace{1cm}
\includegraphics[height=6cm]{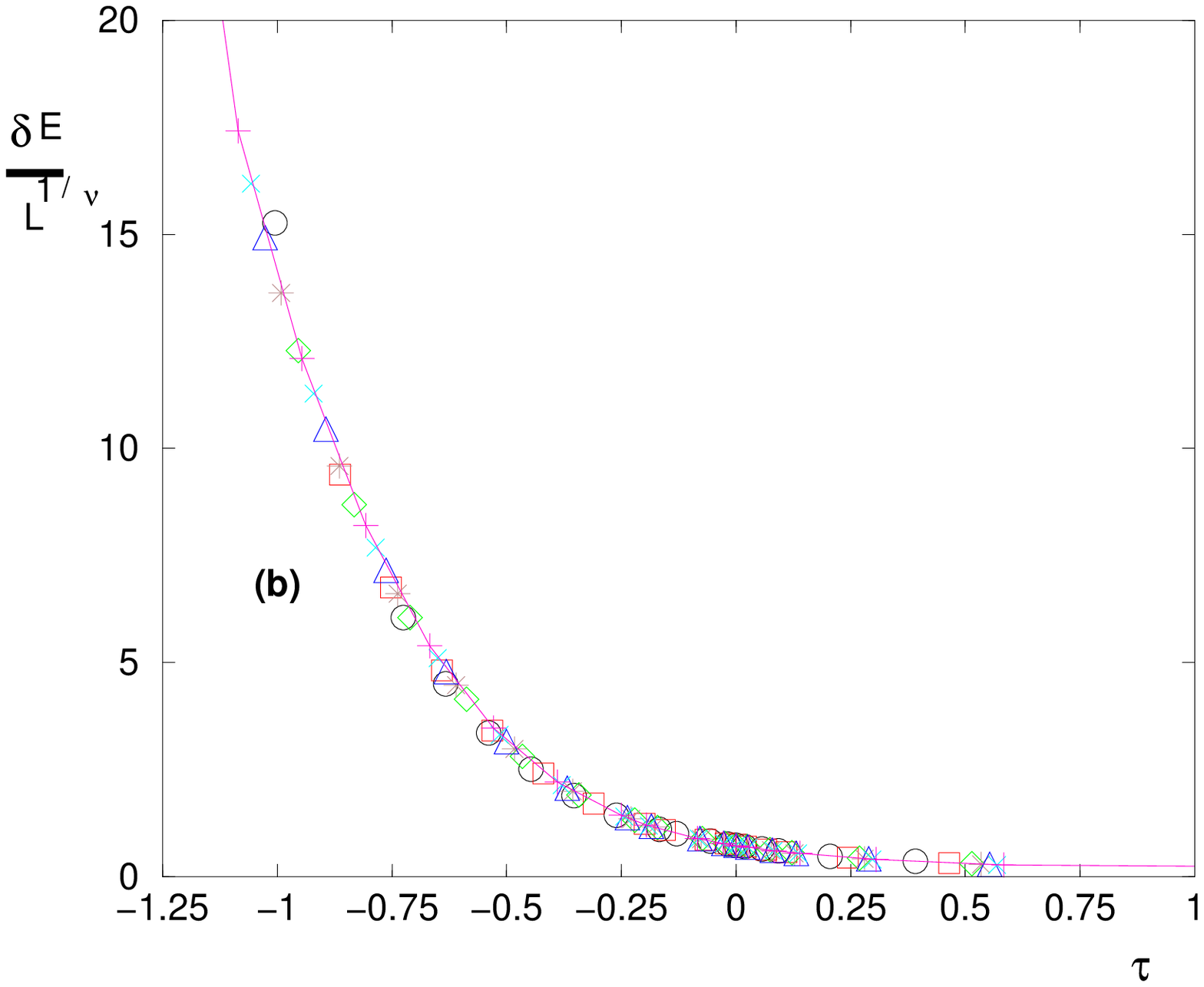}
\caption{
 (a)  Plot of the data of  Fig \ref{fig1} for
$\rho_L(T)=\overline{<R^2_L(T)>}/L$ , as a function of the rescaled
variable $\tau=(T-T_c)L^{1/\nu}$ , see Eq. (\ref{phiscaling}), with
$T_c=0.79$ and $\frac{1}{\nu} \sim 0.475$. Symbols :
$L=30$  $(\bigcirc)$ , $36$ $(\square)$  , $42$ $(\lozenge)$, $48$
$(\triangle)$, $54$ $(\times)$, $60$ $(+)$
(b) Plot of the rescaled energy $\frac{\delta E}{L^{1/\nu}}$ of Eq. 
(\ref{fssenergy}) as a function of the rescaled variable
$\tau=(T-T_c)L^{1/\nu}$, with $T_c=0.79$ and $\frac{1}{\nu} \sim
0.25$. Symbols: $L=12$  $(\bigcirc)$ , $24$ $(\square)$  , $36$ 
$(\lozenge)$, $48$ $(\triangle)$, $54$ $(\times)$, $60$ $(+)$. } 
\label{fig9}
\end{figure}

Fixing the critical temperature at $T_c=T_2=0.79$, we have tried to
make some finite-size scaling analysis
to determine the value of the correlation length exponent $\nu$,
which is related to the specific heat exponent $\alpha$ via the
hyperscaling relation $\nu=2-\alpha$.

We first consider the spatial properties discussed in Section \ref{spatial}.
The best rescaling of the data of Fig.(\ref{fig1} a)
 according to the finite size scaling form
\begin{equation}
\label{phiscaling}
\frac{ \overline{ < R^2_T(L) >}}{L}=\Phi \left( (T-T_c) L^{1/\nu} \right)
\end{equation}
 corresponds to values
in the range $0.45 < \frac{1}{\nu} < 0.5$,
see Fig. (\ref{fig9} a). 

We then consider the averaged energy, with the following
usual finite-size scaling form
\begin{equation}
\label{fssenergy}
\delta E=E_T^{av}(L)-L e_{ann} = L^{1/\nu} \Psi \left( (T-T_c)
L^{1/\nu} \right) 
\end{equation}
The best rescaling (see Fig. (\ref{fig9} b)) correspond to
$0.23<\frac{1}{\nu} < 0.27$. 

These two values (as the previous ones found in
Ref. \cite{Der_Gol,Ki_Br_Mo}) are not compatible. As a consequence, we
cannot give a reasonable quantitative estimate of the critical behavior.
In this respect, we recall that an alternative to the usual power-law
behavior $\xi(T) \sim (T_c-T)^{-\nu}$, has been proposed in 
ref \cite{DPcritidroplet}, namely  $\ln \xi(T) \sim \ln^2
(T_c-T)+...$. Unfortunately, our numerical data are not able to 
test this alternative form without any knowledge of subleading
divergent terms.

\section{ Conclusion}

\label{conclusion}

In this paper, we have presented numerical results
for the directed polymer in a $1+3$ dimensional
 random medium. Our data concerning the wandering $R_T^2(L)$ of the
end-point and the free energy $ F_T(L)$ point towards
$0.76 < T_c \leq T_2=0.79$, so our conclusion is that the critical
temperature $T_c$ is equal or very close to the upper bound $T_2$
derived by Derrida and coworkers. We have also presented results for
histograms of free-energy, energy and entropy as $T$ varies. For $T
\gg T_c$, we obtain that the free-energy distribution is Gaussian. For
$T \ll T_c$, the free-energy distribution coincides with the ground
state energy distribution, in agreement with the zero-temperature
fixed point picture, and the entropy fluctuations of order $\Delta S
\sim L^{1/2}$ follow a Gaussian distribution, in agreement with the
droplet predictions. However, from our various data, we cannot give a
reasonable quantitative estimate of the divergence of the correlation
length $\xi(T)$ as $T \to T_c^-$. 

\section{Acknowledgments}

We thank H. Spohn for pointing out ref \cite{birkner} to us.


\end{document}